\documentstyle[preprint,aps]{revtex} 
\begin{document} \draft

\hfuzz 15pt

\title{Points, Walls and  Loops\\ in Resonant
Oscillatory Media}

\author{Tetsuya Kawagishi, Tsuyosi Mizuguchi and Masaki Sano}

\address{Research Institute of Electrical Communication \\ Tohoku
University, Sendai 980, Japan}

\maketitle


\begin{abstract}

 In an experiment of oscillatory media, domains and walls are formed
under  the parametric resonance with a frequency double the natural
one. In this bi-stable system, 
nonequilibrium transition from Ising wall to Bloch wall consistent
with prediction is confirmed experimentally. The Bloch wall moves in
the direction determined by its chirality with a constant speed. As a new
type of moving structure in two-dimension, a traveling
 loop consisting of two walls and Neel points  is observed.

\end{abstract}



Oscillatory media spontaneously form in a wide variety of systems as
they are driven away from equilibrium. Examples are seen in many
different fields\cite{physics,chemistry,biology}. In nonlinear
oscillators having a small number of degrees of freedom, one of the
central problem is to clarify synchronization and resonance. However,
inspite of the development of theory and experiment,  relatively little
is known about  the response to the external periodic forcing for
spatially coupled oscillators.  The simplest but unresolved case is
parametric resonance of the system where nontrivial  domain walls are
predicted to exist\cite{coullet}.  In the parametric resonance, when the
external forcing has the frequency double the natural frequency, two
locked states are possible.   The domain wall appears between two
different locked states. In this system,
sustained motions of walls are predicted including translational
motion of wall, and oscillating walls due to the non-variational
effect\cite{coullet,Mizuguchi}  even if  two domains are symmetric.
This is the crucial difference from domain walls in equilibrium systems
whose dynamics is simply a relaxational process governed by a free
energy.  We present an experimental investigation of parametrically
forced oscillatory media. A structural transition of domain walls
associated with chirality breaking is elucidated. Furthermore, traveling
loops consisting of two types Bloch walls and Neel points are observed.

Oscillatory media are realized by a convective cellular structure called
Oscillating Grid Pattern(OGP)\cite{sano} in liquid crystal
convection\cite{deGennes}.   We use a nematic liquid crystal, 4-
methoxybezyliden-4'-butylaniline (MBBA) doped with 0.01wt\% of ion
impurity, tetra-n-butylammonium bromide, to control the electrical
conductivity. This liquid crystal is filled in a cell (\( 2cm \times 2cm
\times 50 \mu m \))  sandwiched between transparent
electrodes.
The temperature of the cell is controlled to 25$\pm$0.01$^\circ$C.
Applying an alternative (AC) voltage  of $V \sim$60[V] with frequency
$\omega/2\pi \sim$900[Hz] , we obtain stationary Grid Pattern(GP). It
gives rise a two-dimensional lattice of about 400$\times$400
rectangle convective cells.
Nematic director is stationally since the relaxation time is much
longer (about 0.2sec) than the period of external AC field.\cite{deGennes}
 It is visualized by a shadowgraph under the
microscope with poralized light. The shadowgraphic image intensity is
directly related to the orientation of molecules of
nematics. In this pattern sinks and sources form the
centered rectangle net as shown in Fig.\ \ref{GP}(a).
Slight increase of voltage causes the oscillatory instability of
the Grid Pattern with natural frequency
$\omega_0$; $\omega_0 / 2 \pi  \sim $1.3[Hz] which is independent of
the external frequency $\omega$.
We call this pattern OGP, from the fact that the positions of sinks
and sources move oscillatingly. This is a good candidate of
two-dimensionally distributed oscillatory media.

 To observe behavior of a
parametric resonance of the oscillator lattice, we modulate the AC
voltage with nearly double the natural frequency,  $V_m = 2\sqrt{2}V
(1+r \cos{\omega_e t})cos{\omega t}$, where $r$ is the modulation
ratio and $\omega_e = 2 \omega_0 + \Delta \omega$ and the detuning
$\Delta \omega$ is small. Typical patterns observed in a phase locked
state under
parametric resonance are shown in
Figs.\ref{GP}(b) and (c). The dark lines
  are interfaces between two phase locked states. Across the
dark line  the phase of the oscillation jumps by \( \pi \) as will be
shown later. The interfaces are  walls in dynamical systems.
Unlike in equilibrium systems, these walls can
exhibit transition from stationary to propagating ones
by varying the
control  parameters; $r$ and $\omega_e$. We show a phase diagram in
Fig.\ref{phase-dia}(a) with varying them. Here the AC voltage for
convection was fixed at $\omega/ 2 \pi =$928[Hz], $V =$66.3[V], which
corresponds to a little above the onset of the oscillatory instability of
GP ($\mu \equiv (V-V_c)/V_c=0.009 \pm 0.001 $ where $V_c$ is the
critical voltage for Hopf bifurcation).  In the right half part of phase
locked region, the wall exhibits stationary spatial periodic patterns
(stripe) as shown in Fig.\ref{GP}(c). Recently this periodic pattern of
walls (stripe) are  studied theoretically\cite{Riecke,Walgaraef,coullet93}.
As we decrease the modulation
frequency, the
wavelength of the pattern increases,
 and finally results in isolated walls in the left half part of
the phase locked region as shown in Fig.\ref{GP}(b).
Here we focus on this region in which
stationary or propagating isolated walls are observed.
To elucidate the transition from stationary to propagating walls, we
measure the velocity of moving wall as a function of modulation
frequency $\Delta \omega$ (Fig.\ref{phase-dia}(b)) with fixing $r=0.13$
 (Fig.\ref{phase-dia}(b)).
 It indicates that the
transition is a second order consistent with prediction\cite{coullet}.

It was predicted that the spontaneous breaking of chirality is
responsible for the transition from stationary to moving walls in
nonequilibrium systems\cite{coullet}. The stationary one is called
Ising walls and the moving one is Bloch walls, relying on the analogy to
an anisotropic X-Y model \cite{Magnet}. If the observed stationary
interfaces are Ising walls, the amplitude of oscillation must vanish
where the phase jumps. Furthermore if moving interfaces are Bloch
walls, the amplitude does not vanish at the core, but two domains are
connected by rotating the vector of the complex order parameter of the
oscillation. Now chirality of the wall is defined by the direction of this
rotation, {\it i.e.}, right-handed or left-handed\cite{coullet}. The
moving direction of the wall is determined by the chirality. Hence, one
can judge wall types by its amplitude at the core and chirality.

In order to clarify the structural transition of walls we perform
analysis as followings.
The image intensity ${\cal G}(x,y,t)$ was digitized with a
resolution of 640\( \times \)480 pixels and 256 grey scale levels at
frequency of 15Hz. Here we choose $x$-axis to be parallel to the
direction of the
alignment of nematics.

Under the parametric resonance, OGP has temporal frequency
$\omega_0$ (1.3Hz), and lattice wavenumber $ ( k_x , k_y)
 = (2\pi/63,2\pi/105)(\mu m^{-1}) $.
Thus at the lowest order one can expand ${\cal G} ( x,y,t )$ as follows,
\begin{eqnarray} {\cal G} ( x,y,t )  = [ 1 + a\exp i ( \omega_0 t + \psi ) ]
\exp( ik_x x+ ik_y y ) + c.c. + h.o.t.,  \label{oscig} \end{eqnarray} where
\( a \) represents the amplitude and \( \psi \)  the phase of oscillation
mode, and  {\it h.o.t.} denotes higher order terms.  These amplitude and
phase are slowly varying functions in space and time, compared with
lattice wave number and natural frequency. In order to obtain only the
slow variations in the Eq.(1), we use a broad band pass filter centered
at $\omega_0$ in frequency and $(k_x,k_y)$ in wavenumber.  Thus we
obtain filtered signal; $a \exp(i\psi)$, which is the complex order
parameter of our system.

Although the oscillation and interfaces are two-dimensional
phenomena, to show temporal evolutions of interfaces evidently at
first we take a one-dimensional section of oscillators  intersecting a
wall perpendicularly. Figure\ \ref{Ising} shows the result about a
stationary wall. In Figure\ \ref{Ising}(a) the solid line
represents  the spatial variation of the phase, and there exists a jump
about \( \pi \) in its center. At the same point the amplitude (dotted
line) falls almost to zero.
A spatio-temporal plot of amplitude profile in Figure\ \ref{Ising}(b)
shows evidently that
the wall is stationary. Therefore we conclude that stationary walls are
Ising walls. On the other hand, in Fig.\ \ref{Bloch} we show the result
about the moving wall.   The amplitude at the walls are relatively small
but does not vanish (dotted line in Fig.\ \ref{Bloch}(a)). Phase gradually
changes by $\pi$ at the center of wall (solid line in Fig.\
\ref{Bloch}(a)) which slope is less steeper than the case of Ising wall.
Notice that the wall is moving leftward in spatio-temporal plot of phase
profile (Fig.\ \ref{Bloch}(b)).
Consequently it is concluded that moving walls are Bloch wall. As in an
easy-axis ferromagnets, there are two kinds of Bloch walls in this
system. The phase decreases from left domain to right one in Fig.\
\ref{Bloch}.  The other type of Bloch walls exhibiting
rightward motion are also observed,  which connects the same domains with
increasing the phase.   Hence, the motion of Bloch wall is determined
by its chirality.

 Let us discuss two-dimensional structures and behavior.
In two dimension, Bloch walls can form  closed loops. The simplest loop
is the one consisting of only a single type of Bloch wall
which expands invading
outer region or shrinks invading inner region depending on its chirality.
(see Fig. \ref{loop}(a),(b))
However interesting motion is expected if the loop consists of two
different types of Bloch walls jointed at two points. (Fig. \ref{loop}(c))
This point is called Neel point or Connecting
Point(CP)\cite{Mizuguchi,Frish}.  Some interesting phenomena are
expected because of  the interplay between two types of Bloch walls
jointed by CP. Among them we report a new type of moving structure  in
two-dimensional space; a translational motion of a loop. Figure\
\ref{CP}(a) shows a snapshot of a moving loop exhibiting translational
motion to the right. The loop travels persistently to one direction
 until it collides with other walls or loops. We analyzed the two-
dimensionally distributed oscillators near the upper end of the loop.
Figure\ \ref{CP}(b) shows the spatial phase variation. It is seen that
two different Bloch walls are connected at a kind of branch cut of a
$2\pi$ jump.  In Fig.\ \ref{CP}, the left domain and the right domain are
the same since they are connected out side the loop. The branch cut is
resulted from this fact. Notice that the center domain is connected
with right one by increasing the phase and with left one by decreasing.
These increasing and decreasing correspond to left- and right-handed
rotation of the vector of complex order parameter. Their directions of
motions are opposite to each other, {\it i.e.}  one invades outward of a
loop and the other does inward. Hence, again the motion of Bloch wall is
determined uniquely by its chirality.  Considering the topology of the
loop,  two CPs must exist at the upper and lower ends of the loop which
correspond to the singular points of the branch cut. The amplitude of
Bloch wall does not vanish, nevertheless at only CP it must vanish due
to topological constraint. We plot the cross section of the amplitude
profile by crossing interfaces twice at a Bloch wall and CP in Fig.\
\ref{CP}(c). Evidently Fig.\ \ref{CP}(c) shows that the amplitude is
relatively small at Bloch wall, and  zero at CP. This confirms that
 the traveling
loop consists of two different types of Bloch walls and two CPs as
schematically illustrated in Fig.\ref{loop}(c).

In the traveling loops, CPs move with the Bloch walls.
This is in contrast with usual spiral patterns in which the core
(CP) remains unmoved.\cite{Meyer,Frish,Nasuno94}
A possible
reason is following. In the present experiment, the system is fully non-
variational because the system consisting of limit cycle oscillators.
In fact, the traveling loop is observed numerically  in  CGL equation with
parametric forcing term by appropriate choice of
 the linear dispersion coefficient
and the detuning parameter\cite{Mizuguchi94}.
Recently observed traveling spots in a model of reaction diffusion equation
\cite{Krischer}
 seems very similar to the present traveling loops.
Therefore we believe
that the phenomena are generic in (resonant) oscillatory media.
The mechanism of the motion of CP may be related to core meandering in
spiral patterns. It is an open problem in resonant system. The motion of
the loops and connecting point will open an interesting question
about dynamics of
interfaces; walls, lines and points, appearing in higher dimensional
space for nonequilibrium systems\cite{Kleman}.

The authors wish to thank Y.Sawada, Y.Kuramoto, S.Sasa,
and H.Sakaguchi for valuable discussions, and to H.Kokubo
for helpful suggestion on experiment.



\begin{figure}
\caption{ The shadowgraphic image of the Grid Pattern. The length of bars
corresponds to 100$\mu m$. (a)A snapshot
of oscillating Grid Pattern(OGP). (b)OGP under the parametric
resonance. Dark lines are domain wall between two different locked
states. (c)same as (b) but showing stripe pattern.}{ \label{GP}}
\end{figure}

\begin{figure}\caption{(a) The phase diagram of the
parametric resonance as a function of modulating ratio $r$ and
frequency $\omega_e$. (b) Traveling velocity of Ising and Bloch walls as a
function of frequency detuning $\Delta \omega$
with fixing $r = 0.13$.}{\label{phase-dia}}
\end{figure}

\begin{figure}
\caption{ Stationary interface for one-dimensional
oscillators along a line intersecting the wall.
(a)the spatial variation of the amplitude (dotted line) and the
phase (solid line) (b)space time evolution of the the amplitude profile.
Black (white) pixels correspond to a small (large) amplitude of
oscillation.}{\label{Ising}}
\end{figure}

\begin{figure}\caption{ Moving interface for one-dimensional
oscillators along a line intersecting the wall.
(a)the spatial variation of the amplitude (dotted
line) and the phase (solid line) (b)space time evolution of the phase
variation. The grey sale corresponds to white for 0 and black for
$2\pi$.}{\label{Bloch}}
\end{figure}

\begin{figure}\caption{Schematic illustration of possible types of loops.
 (a) An expanding loop consists of a Bloch wall (+).
(b) A shrinking loop consists of a Bloch wall ($-$). (c) A traveling loop
consists of two Bloch walls (+ and $-$) and Neel points(CPs).
}{
\label{loop}}
\end{figure}

\begin{figure}\caption{ Experimental data of a traveling loop
 (a) Phase distribution of a snapshot
of a loop exhibiting translational
motion to rightward. The bar corresponds to 100$\mu m$.
(b) a spatial variation of the phase around upper end
of the loop. (c) a cross
section of amplitude profile which crosses a Bloch
wall (B) and a CP. The amplitude vanishes at the CP.}{
\label{CP}}
\end{figure}

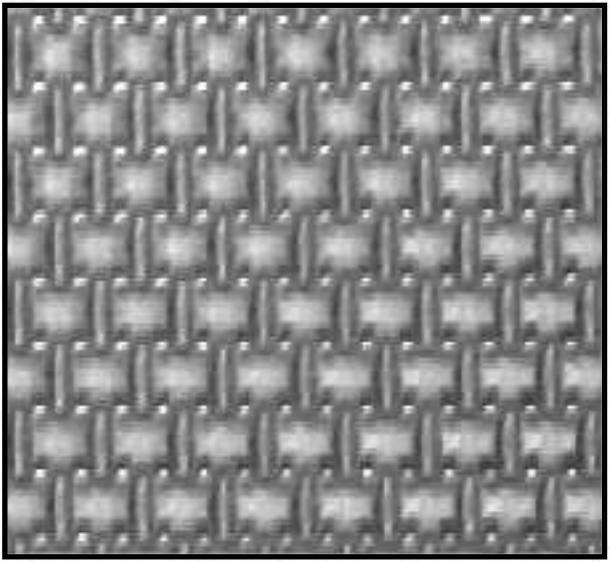
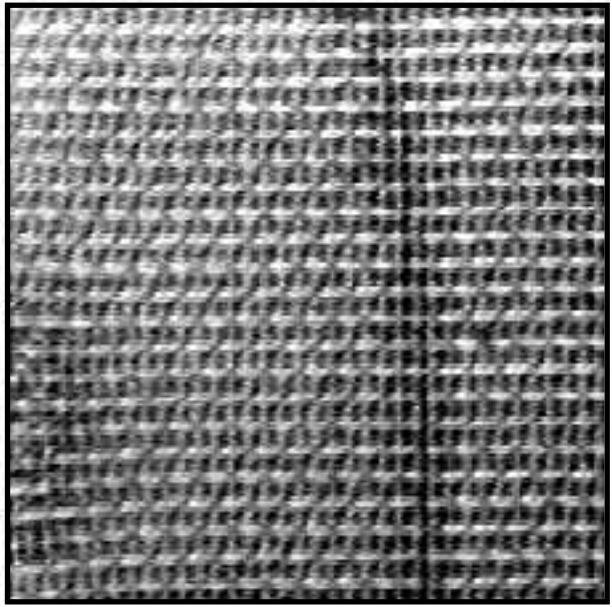
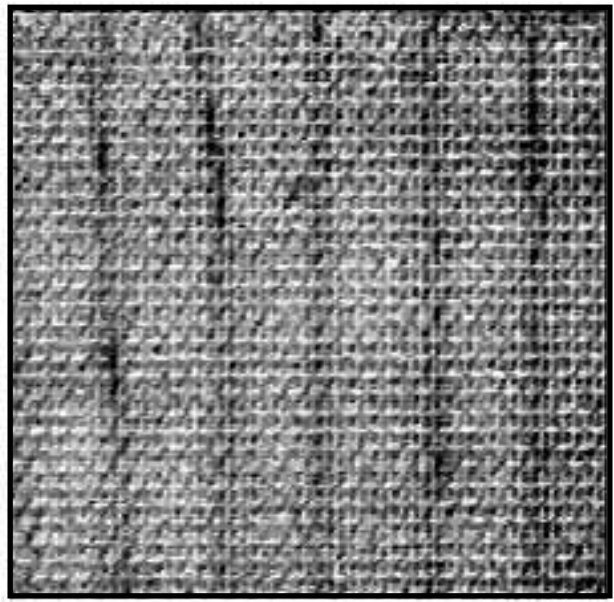
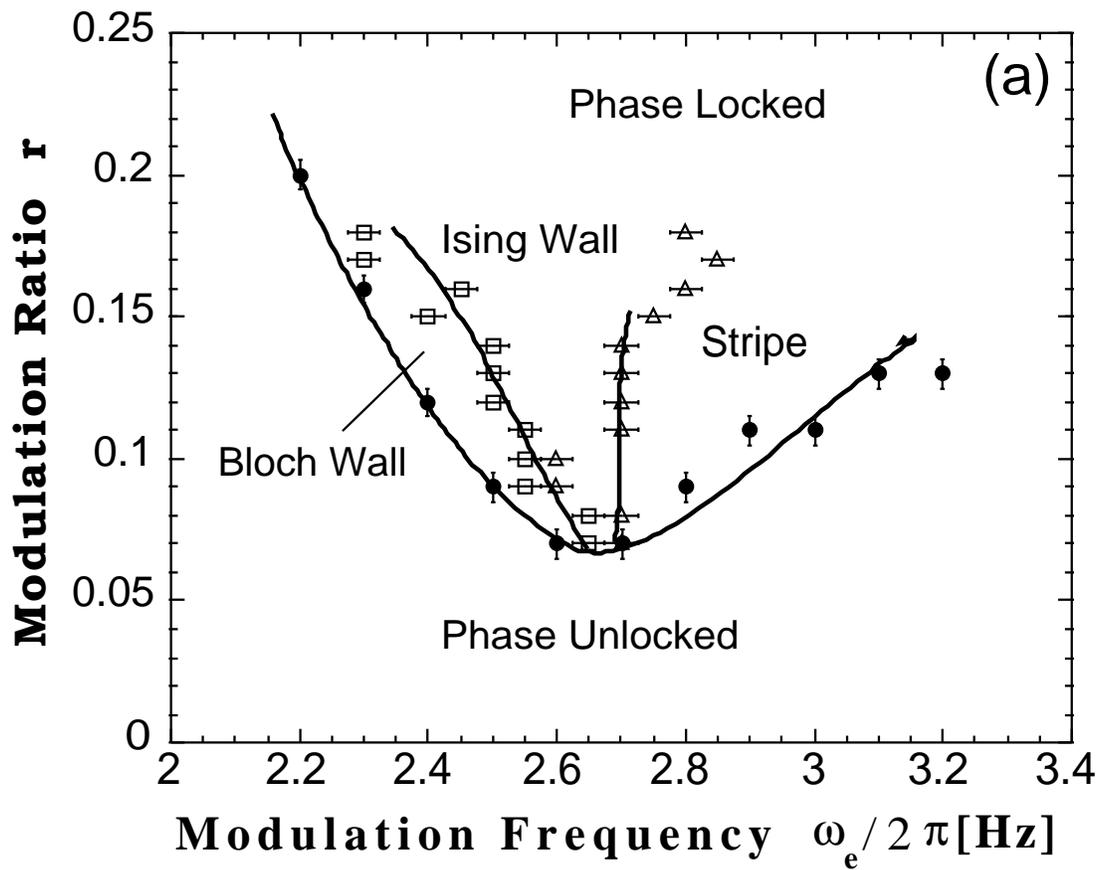
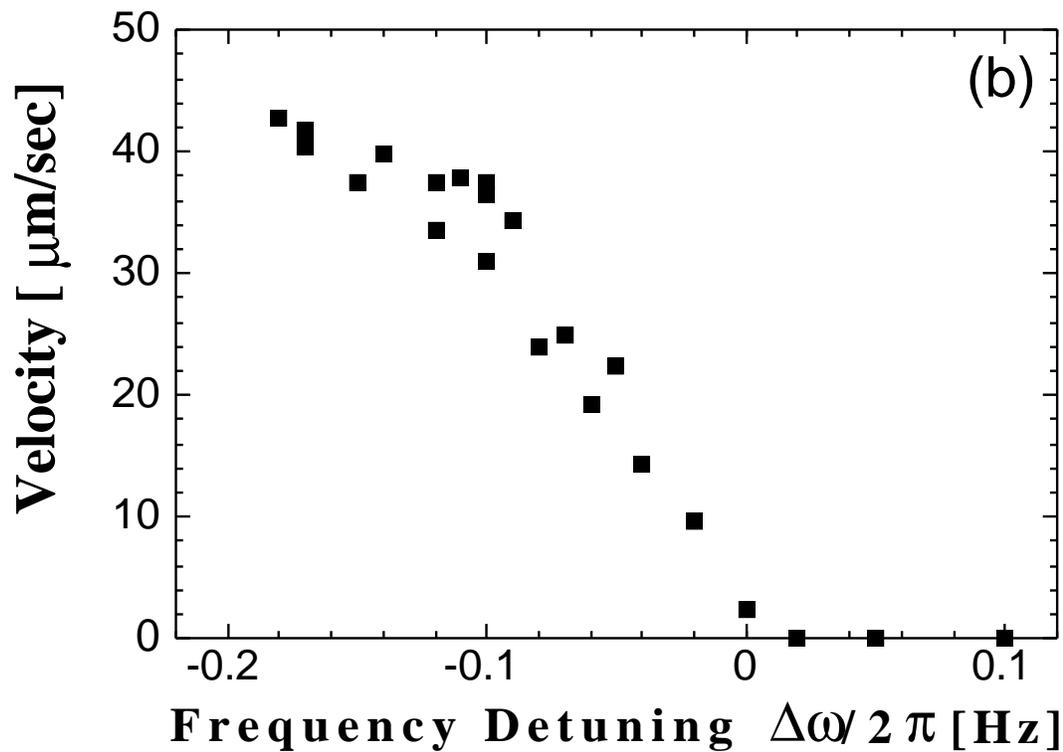
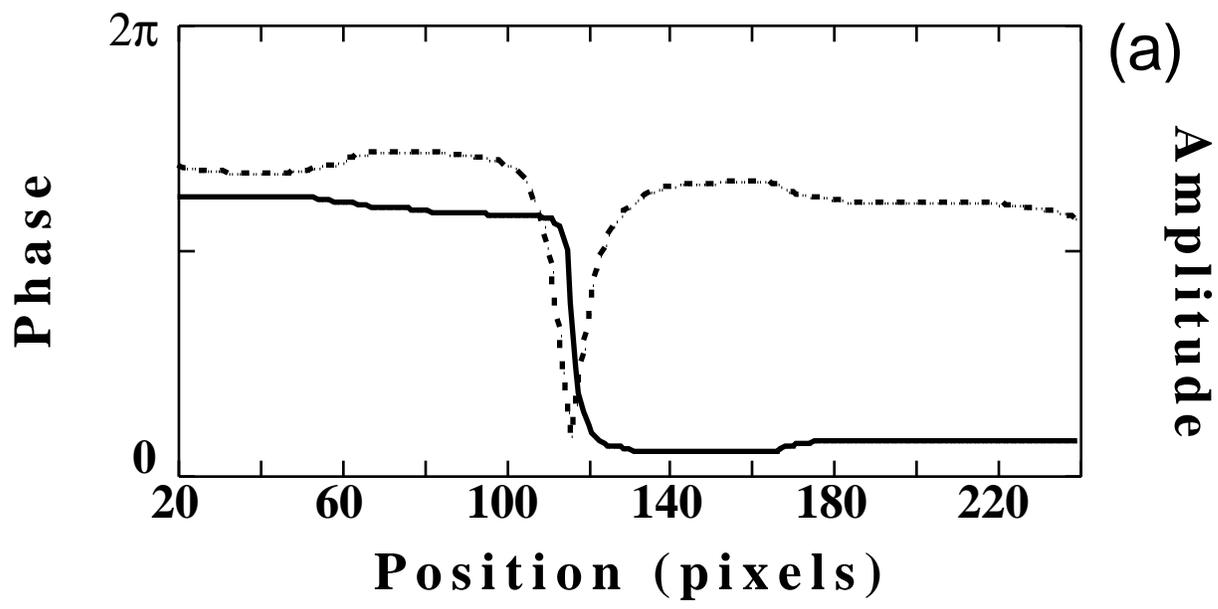
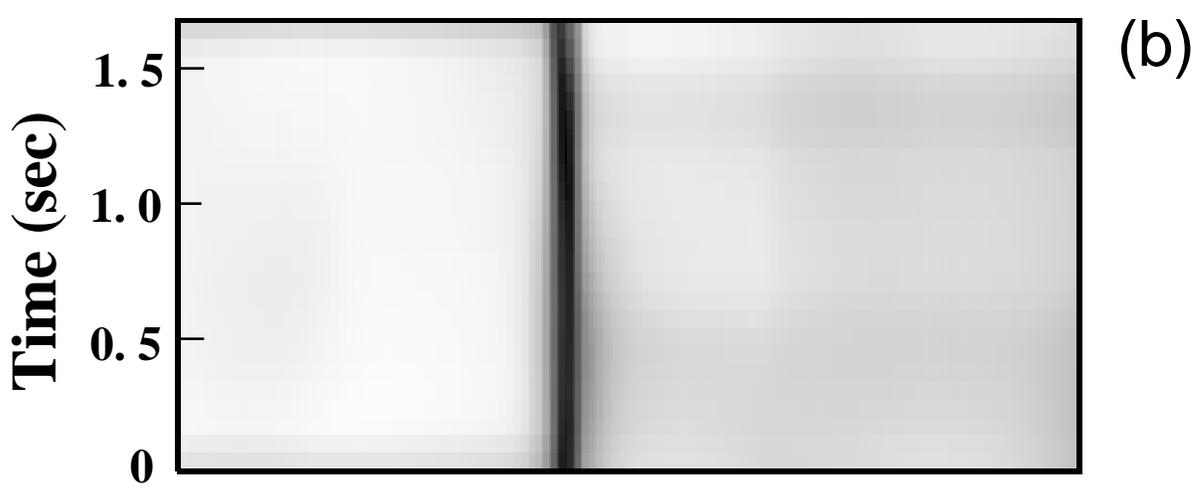
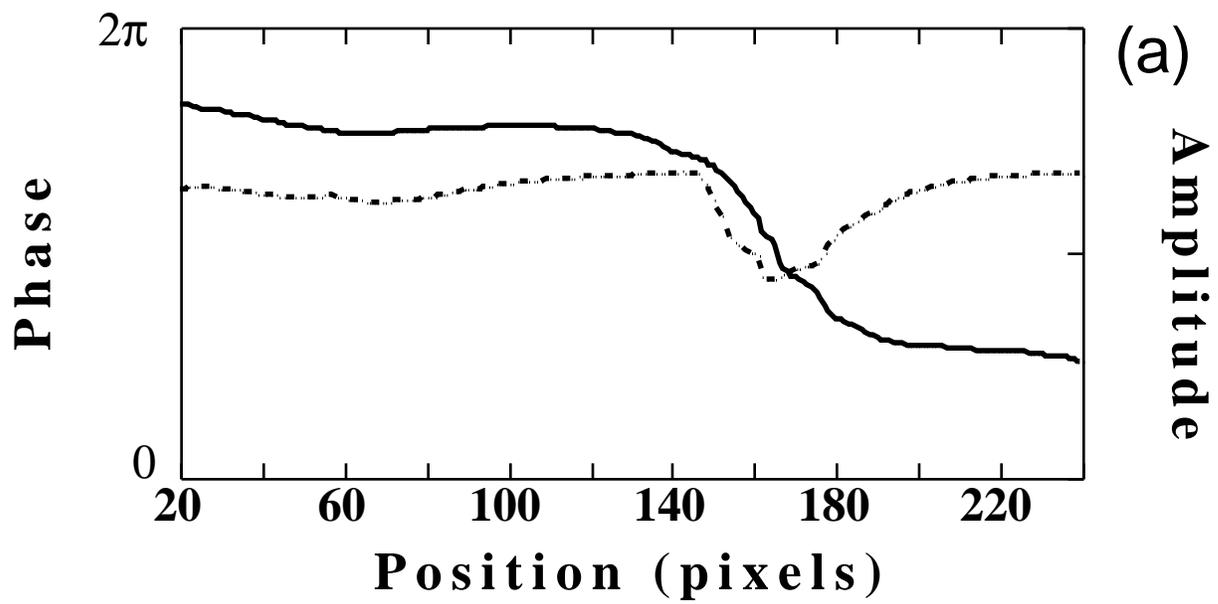
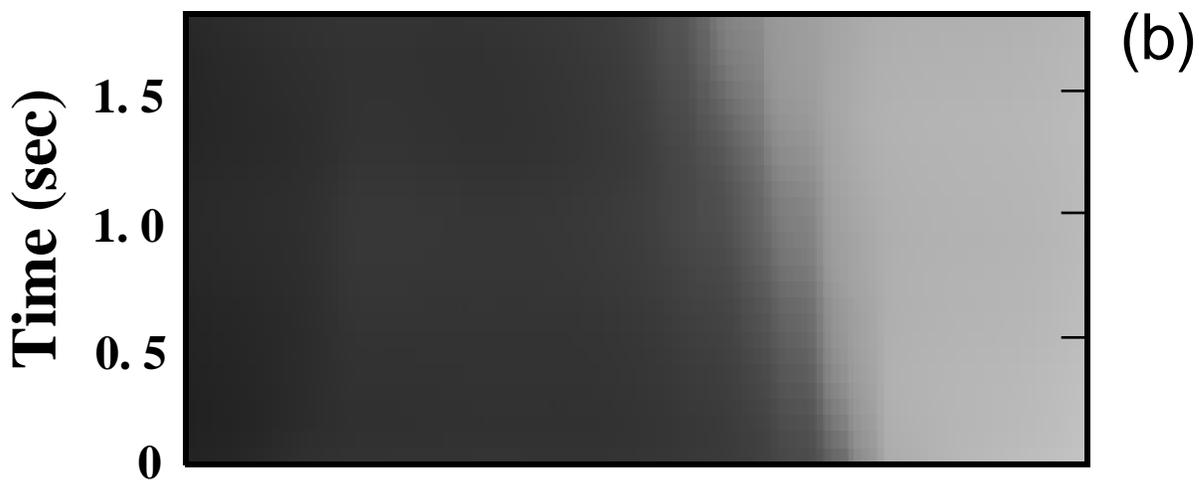
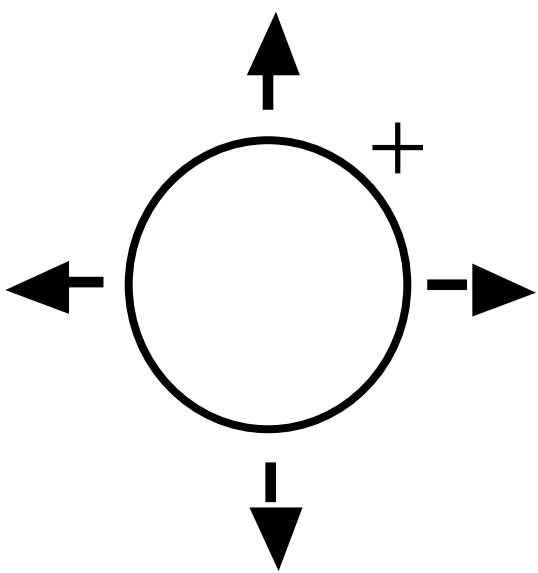
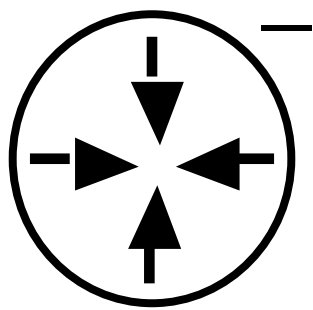
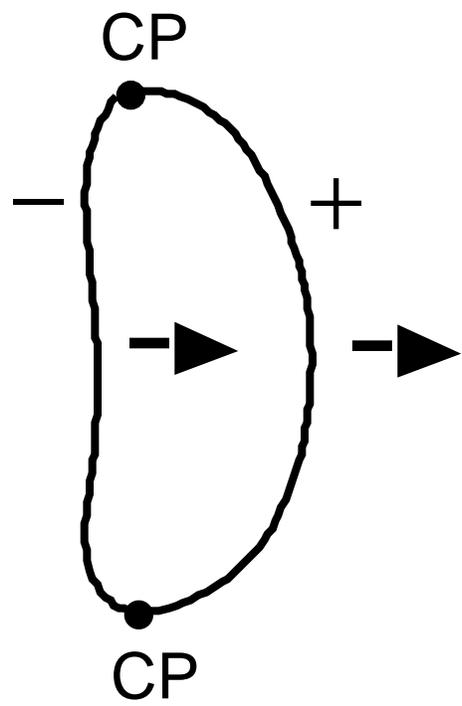
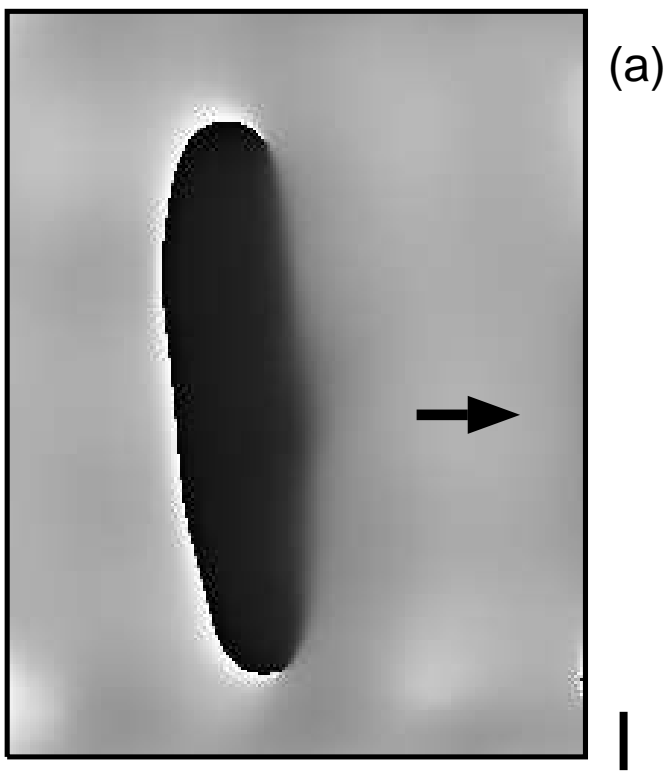
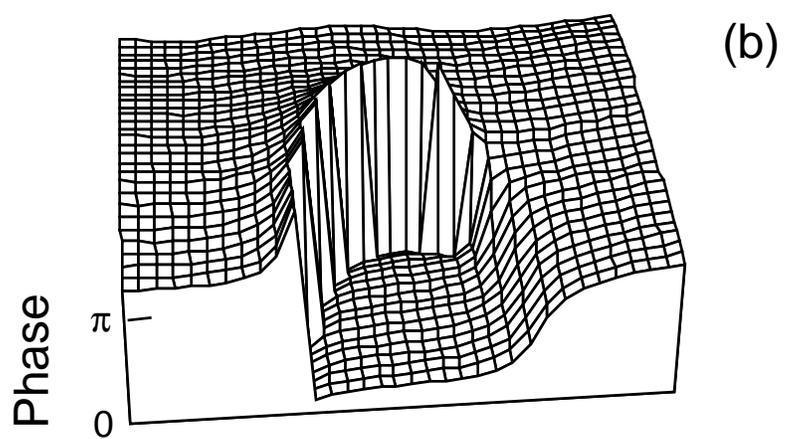
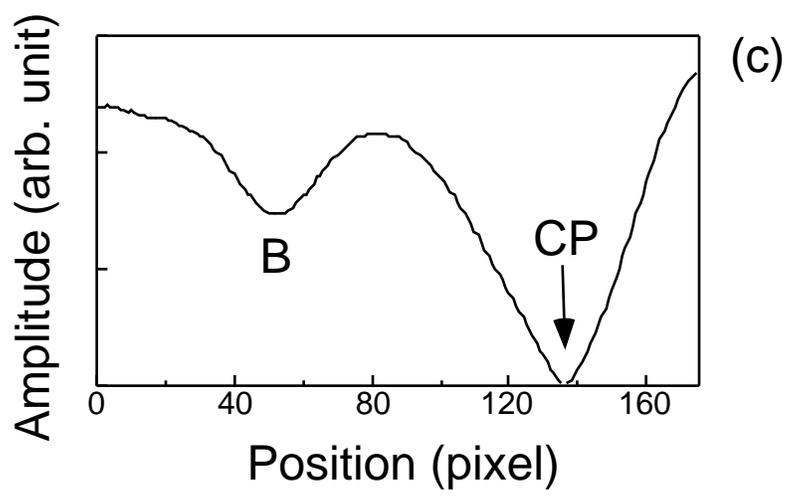
\end{document}